# Spatial correlation of the thermally generated electromagnetic field in layered media

Vahid Hatamipour and Mathieu Francoeur[†]

*Radiative Energy Transfer Lab, Department of Mechanical Engineering, University of Utah, Salt Lake City, UT 84112, USA*

**ABSTRACT**

A general formulation for the cross-spectral density tensor enabling calculation of the spatial correlation of the thermally generated electromagnetic field in layered media is derived. The formulation is based on fluctuational electrodynamics, and is thus applicable in the near and far field of heat sources. The resulting cross-spectral density tensor is written in terms of a single integration over the parallel wavevector, as the angular integrations leading to numerical instability are evaluated analytically. Using this formulation, the spatial correlation length in the near field of a film made of silicon carbide (SiC) supporting surface phonon-polaritons (SPhPs) in the infrared is analyzed. It is shown that the spatial correlation length of a SiC heat source suspended in vacuum decreases substantially by decreasing its thickness owing to SPhP coupling. In the limit of a 10-nm-thick SiC film, the spatial correlation length is similar to that of a blackbody. The results also reveal that it is possible to control the spatial coherence of a thin SiC heat source via dielectric and metallic substrates, respectively allowing and preventing SPhP coupling. This suggests that active modulation of thermal emission via thin films supporting surface polaritons in the infrared is possible by using a phase change material substrate such as vanadium dioxide.

[†] Corresponding author. Tel.: + 1 801 581 5721
Email addresses: mfrancoeur@mech.utah.edu, vahid.hatami@utah.edu

# I. INTRODUCTION

A heat source emits both propagating and evanescent waves. Planck's blackbody theory, valid when all length scales are much larger than the wavelength, only takes into account propagating waves [1]. Evanescent waves, which include frustrated modes and surface polaritons, decay within a distance of about a wavelength or less normal to the surface a heat source and play a key role in near field heat exchange [2]. For instance, when two heat sources are separated by a sub-wavelength distance, radiative heat transfer can exceed the blackbody limit owing to tunneling of evanescent waves [3-7].

Evanescent waves can also greatly modify the coherence properties of heat sources in the far field. Heat sources are typically incoherent, implying that thermal emission is broadband and quasi-isotropic. Carminati and Greffet [8] however showed that the spatial correlation length in the near field of a heat source made of silicon carbide (SiC) supporting surface phonon-polaritons (SPhPs) in the infrared can far exceed that of a blackbody owing to the long propagation length of SPhPs. Greffet et al. [9] later experimentally demonstrated that a large spatial correlation length in the near field of a SiC heat source leads to highly directional thermal emission at specific wavelengths by coupling SPhPs to the far field via a grating. Coherent thermal emission mediated by excitation of surface polaritons via gratings was also reported in Refs. [10-17].

The spatial correlation of the electric field near a heat source is quantified by the cross-spectral density tensor [18] within the framework of fluctuational electrodynamics [19], where a fluctuating current representing thermal emission is added to Maxwell's equations. Expressions for the cross-spectral density tensor in the near field of bulk [8,20] and film [21,22] heat sources have been derived. These expressions, however, are limited to specific geometries and are written in terms of angular integrals that are difficult to solve numerically. There is a lack of a general formalism enabling calculation of the cross-spectral density tensor in layered media. The objective of this paper is, therefore, to provide an exact formalism for calculating the cross-spectral density tensor in layered media via fluctuational electrodynamics. The proposed formalism is general, in the sense that it is applicable both in the near- and far-field of heat sources. In addition, the angular integrals are solved analytically, thus greatly simplifying the calculation of the cross-spectral density tensor. By applying the formalism to a film heat source made of SiC, it is shown that its spatial coherence can be modulated by changing the film thickness and the substrate onto which the film is coated.

The rest of the paper is organized as follows. A general expression for the cross-spectral density tensor in layered media is first derived starting from fluctuating electrodynamics and using dyadic Green's functions (DGFs). Next, the impact of SPhP coupling in a SiC film heat source suspended in vacuum and coated on dielectric and metallic substrates is analyzed. Concluding remarks are provided in section IV. Finally, the Appendix provides explicit expressions for the nine components of the cross-spectral density tensor in layered media.

## II. CROSS-SPECTRAL DENSITY TENSOR IN LAYERED MEDIA

The spatial correlation of the electromagnetic field is characterized by the cross-spectral density tensor defined as [18]:

$$\bar{\bar{\mathbf{W}}}(\mathbf{r}_1,\mathbf{r}_2,\omega)\delta(\omega-\omega') = \langle \mathbf{E}(\mathbf{r}_1,\omega) \otimes \mathbf{E}(\mathbf{r}_2,\omega') \rangle \qquad (1)$$

where $\otimes$ denotes the outer product and the angled brackets $\langle \ \rangle$ represent an ensemble average. Eq. (1) provides a measure of the spatial correlation of the electric field between locations $\mathbf{r}_1$ and $\mathbf{r}_2$ ($\mathbf{r} = x\hat{\mathbf{x}} + y\hat{\mathbf{y}} + z\hat{\mathbf{z}}$) for a frequency $\omega$ [8]. Under the assumption that the materials are non-magnetic, the thermally generated electric field, derived from fluctuational electrodynamics [19], is given by:

$$\mathbf{E}(\mathbf{r},\omega) = i\omega\mu_0 \int_V \bar{\bar{\mathbf{G}}}(\mathbf{r},\mathbf{r}',\omega) \cdot \mathbf{J}^{fl}(\mathbf{r}',\omega) d^3\mathbf{r}' \tag{2}$$

where $\mu_0$ is the permeability of vacuum, $\bar{\bar{\mathbf{G}}}(\mathbf{r},\mathbf{r}',\omega)$ is the DGF relating the electric field at a point $\mathbf{r}$ to a source located at $\mathbf{r}'$, and $\mathbf{J}^{fl}(\mathbf{r}',\omega)$ is the fluctuating current acting as the thermal radiation source. The spatial correlation of the thermally generated electric field at locations $\mathbf{r}_1$ and $\mathbf{r}_2$ written in terms of the fluctuating current and DGFs is obtained by inserting Eq. (2) in Eq. (1):

$$\bar{\bar{\mathbf{W}}}(\mathbf{r}_1,\mathbf{r}_2,\omega)\delta(\omega-\omega') = \omega^2\mu_0^2 \int_V d^3\mathbf{r}' \int_V d^3\mathbf{r}'' \bar{\bar{\mathbf{G}}}(\mathbf{r}_1,\mathbf{r}',\omega)\langle \mathbf{J}^{fl}(\mathbf{r}',\omega) \otimes \mathbf{J}^{fl}(\mathbf{r}'',\omega')\rangle \bar{\bar{\mathbf{G}}}^{\dagger}(\mathbf{r}_2,\mathbf{r}'',\omega') \tag{3}$$

where the superscript † denotes conjugate transpose. The ensemble average of the spatial correlation function of the fluctuating current is related to the local temperature of a heat source via the fluctuation-dissipation theorem [19]:

$$\langle \mathbf{J}^{fl}(\mathbf{r}',\omega) \otimes \mathbf{J}^{fl}(\mathbf{r}'',\omega')\rangle = \frac{4\omega\varepsilon_0\varepsilon''(\omega)}{\pi}\Theta(\omega,T)\delta(\mathbf{r}'-\mathbf{r}'')\delta(\omega-\omega')\bar{\bar{\mathbf{I}}} \tag{4}$$

where $\varepsilon_0$ is the permittivity of vacuum, $\varepsilon''(\omega)$ is the imaginary part of the dielectric function of the heat source, $\bar{\bar{\mathbf{I}}}$ is the unit dyadic, and $\Theta(\omega,T) = \hbar\omega/[\exp(\hbar\omega/k_B T)-1]$ is the energy of an electromagnetic state with $k_B$ as the Boltzmann constant and $\hbar$ as the reduced Plank constant. The application of the fluctuation-dissipation theorem to Eq. (3) leads to the following simplified cross-spectral density tensor:

$$\bar{\bar{\mathbf{W}}}(\mathbf{r}_1,\mathbf{r}_2,\omega) = \frac{4\omega^3 \mu_0^2 \varepsilon_0 \varepsilon''(\omega)}{\pi} \Theta(\omega,T) \int_V d^3\mathbf{r}' \bar{\bar{\mathbf{G}}}(\mathbf{r}_1,\mathbf{r}',\omega) \bar{\bar{\mathbf{G}}}^\dagger(\mathbf{r}_2,\mathbf{r}',\omega) \qquad (5)$$

The formulation is hereafter specialized for the layered geometry shown in Fig. 1, where the spatial correlation of the electric field at locations $\mathbf{r}_1$ and $\mathbf{r}_2$ in layer $l$ due to thermal emission by layer $s$ is of interest. For layered media, it is convenient to use a plane wave representation of the DGF [23,24]:

$$\bar{\bar{\mathbf{G}}}(\mathbf{r},\mathbf{r}',\omega) = \int_{-\infty}^{\infty} \frac{d^2\mathbf{K}}{4\pi^2} \bar{\bar{\mathbf{g}}}(\mathbf{K},z,z',\omega) e^{i\mathbf{K}\cdot(\mathbf{R}-\mathbf{R}')} \qquad (6)$$

where $\mathbf{R} = x\hat{\mathbf{x}} + y\hat{\mathbf{y}}$, $\mathbf{K} = k_x\hat{\mathbf{x}} + k_y\hat{\mathbf{y}}$ is the wavevector parallel to the surfaces, $z$ and $z'$ are respectively the $z$-locations in the observation ($l$) and source ($s$) layers, while $\bar{\bar{\mathbf{g}}}$ is the Weyl component of the DGF. Using the plane wave representation of the DGF in Eq. (5), and redefining the integration over the volume as $\int_V d^3\mathbf{r}' = \int_0^{t_s} dz' \int_{-\infty}^{\infty} d^2\mathbf{R}'$, where $t_s$ is the thickness of the source layer $s$, the cross-spectral density tensor is written as:

$$\bar{\bar{\mathbf{W}}}(\mathbf{r}_1,\mathbf{r}_2,\omega) = \frac{4\omega^3 \mu_0^2 \varepsilon_0 \varepsilon''(\omega)}{\pi} \Theta(\omega,T) \int_0^{t_s} dz' \int_{-\infty}^{\infty} d^2\mathbf{R}'$$
$$\times \int_{-\infty}^{\infty} \frac{d^2\mathbf{K}'}{4\pi^2} \bar{\bar{\mathbf{g}}}(\mathbf{K}',z_1,z',\omega) e^{i\mathbf{K}'\cdot(\mathbf{R}_1-\mathbf{R}')} \int_{-\infty}^{\infty} \frac{d^2\mathbf{K}}{4\pi^2} \bar{\bar{\mathbf{g}}}^\dagger(\mathbf{K},z_2,z',\omega) e^{-i\mathbf{K}\cdot(\mathbf{R}_2-\mathbf{R}')} \qquad (7)$$

Rearranging the above expression and applying $\int_{-\infty}^{\infty} d^2\mathbf{R}' e^{i(\mathbf{K}'-\mathbf{K})\cdot\mathbf{R}'} = 4\pi^2 \delta(\mathbf{K}'-\mathbf{K})$, Eq. (7) becomes:

$$\overline{\overline{\mathbf{W}}}(\mathbf{r}_1,\mathbf{r}_2,\omega) = \frac{4\omega^3\mu_0^2\varepsilon_0\varepsilon''(\omega)}{\pi}\Theta(\omega,T)\int_{-\infty}^{\infty}\frac{d^2\mathbf{K}}{4\pi^2}\int_0^{t_s}dz'\overline{\overline{\mathbf{g}}}(\mathbf{K},z_1,z',\omega)\overline{\overline{\mathbf{g}}}^\dagger(\mathbf{K},z_2,z',\omega)e^{i\mathbf{K}\cdot(\mathbf{R}_1-\mathbf{R}_2)} \qquad (8)$$

The infinite two-dimensional integration over the parallel wavevector can be simplified using

$\int_{-\infty}^{\infty}d^2\mathbf{K} = \int_0^{2\pi}d\phi\int_0^{\infty}K\,dK$, where $K$ is the magnitude of $\mathbf{K}$ and $\phi$ is the azimuthal angle in the $x-y$ plane. Inserting this last expression in Eq. (8) in addition to defining the $x$- and $y$-components of the parallel wavevector in terms of the azimuthal angle, $k_x = K\cos(\phi)$ and $k_y = K\sin(\phi)$, the cross-spectral density tensor is simplified as follows:

$$\overline{\overline{\mathbf{W}}}(\mathbf{r}_1,\mathbf{r}_2,\omega) = \frac{4\omega^3\mu_0^2\varepsilon_0\varepsilon''(\omega)}{\pi}\Theta(\omega,T) \\ \times \int_0^{\infty}\frac{KdK}{4\pi^2}\int_0^{2\pi}d\phi\int_0^{t_s}dz'\overline{\overline{\mathbf{g}}}(K,\phi,z_1,z',\omega)\overline{\overline{\mathbf{g}}}^\dagger(K,\phi,z_2,z',\omega)e^{iK(\cos(\phi)\Delta x+\sin(\phi)\Delta y)} \qquad (9)$$

where $\Delta x = x_1 - x_2$ and $\Delta y = y_1 - y_2$.

The next step is to specify the Weyl component of the DGF (see Fig. 2). Defining $\Delta z_1 = z_1 - z_l$ and $\Delta z_2 = z_2 - z_l$, the Weyl component of the DGF at $\mathbf{r}_1$ and its conjugate transpose at $\mathbf{r}_2$ are given by [24]:

$$\overline{\overline{\mathbf{g}}}(K,\phi,z_1,z',\omega) = \frac{i}{2k_{zs}}\begin{pmatrix} \left(A_l^{TE}e^{ik_{zl}\Delta z_1}e^{-ik_{zs}z'} + B_l^{TE}e^{-ik_{zl}\Delta z_1}e^{-ik_{zs}z'}\right)\hat{\mathbf{s}}\otimes\hat{\mathbf{s}} \\ +\left(C_l^{TE}e^{ik_{zl}\Delta z_1}e^{ik_{zs}z'} + D_l^{TE}e^{-ik_{zl}\Delta z_1}e^{ik_{zs}z'}\right)\hat{\mathbf{s}}\otimes\hat{\mathbf{s}} \\ +A_l^{TM}e^{ik_{zl}\Delta z_1}e^{-ik_{zs}z'}\hat{\mathbf{p}}_l^+\otimes\hat{\mathbf{p}}_s^+ + B_l^{TM}e^{-ik_{zl}\Delta z_1}e^{-ik_{zs}z'}\hat{\mathbf{p}}_l^-\otimes\hat{\mathbf{p}}_s^+ \\ +C_l^{TM}e^{ik_{zl}\Delta z_1}e^{ik_{zs}z'}\hat{\mathbf{p}}_l^+\otimes\hat{\mathbf{p}}_s^- + D_l^{TM}e^{-ik_{zl}\Delta z_1}e^{ik_{zs}z'}\hat{\mathbf{p}}_l^-\otimes\hat{\mathbf{p}}_s^- \end{pmatrix} \qquad (10)$$

$$\bar{\bar{g}}^{\dagger}\left(K,\phi,z_2,z',\omega\right)=-\frac{i}{2k_{zs}^*}\begin{pmatrix}\left(A_l^{TE*}e^{-ik_{zl}^*\Delta z_2}e^{ik_{zs}^*z'}+B_l^{TE*}e^{ik_{zl}^*\Delta z_2}e^{ik_{zs}^*z'}\right)\hat{\mathbf{s}}\otimes\hat{\mathbf{s}}\\+\left(C_l^{TE*}e^{-ik_{zl}^*\Delta z_2}e^{-ik_{zs}^*z'}+D_l^{TE*}e^{ik_{zl}^*\Delta z_2}e^{-ik_{zs}^*z'}\right)\hat{\mathbf{s}}\otimes\hat{\mathbf{s}}\\+A_l^{TM*}e^{-ik_{zl}^*\Delta z_2}e^{ik_{zs}^*z'}\hat{\mathbf{p}}_s^+\otimes\hat{\mathbf{p}}_l^++B_l^{TM*}e^{ik_{zl}^*\Delta z_2}e^{ik_{zs}^*z'}\hat{\mathbf{p}}_s^+\otimes\hat{\mathbf{p}}_l^-\\+C_l^{TM*}e^{-ik_{zl}^*\Delta z_2}e^{-ik_{zs}^*z'}\hat{\mathbf{p}}_s^-\otimes\hat{\mathbf{p}}_l^++D_l^{TM*}e^{ik_{zl}^*\Delta z_2}e^{-ik_{zs}^*z'}\hat{\mathbf{p}}_s^-\otimes\hat{\mathbf{p}}_l^-\end{pmatrix} \quad (11)$$

where the superscript * denotes complex conjugate, while $k_{zi}=\sqrt{k_i^2-K^2}$ and $k_i$ are respectively the perpendicular component and the magnitude of the wavevector in layer $i$. The TE- and TM-polarized unit vectors, $\hat{\mathbf{s}}$ and $\hat{\mathbf{p}}$, are defined as [25]:

$$\hat{\mathbf{s}}=\sin(\phi)\hat{\mathbf{x}}-\cos(\phi)\hat{\mathbf{y}} \quad (12)$$

$$\hat{\mathbf{p}}_i^{\pm}=\frac{1}{k_i}\left(K\hat{\mathbf{z}}\mp k_{zi}\left(\cos(\phi)\hat{\mathbf{x}}+\sin(\phi)\hat{\mathbf{y}}\right)\right) \quad (13)$$

In Eqs. (10) and (11), the coefficients $A_l^{\gamma}$ and $B_l^{\gamma}$ are respectively the amplitudes of forward and backward traveling waves, with respect to the $z$-axis, in layer $l$ and polarization state $\gamma$ generated by a source emitting in the forward direction located in layer $s$. The same definition holds for the coefficients $C_l^{\gamma}$ (forward traveling) and $D_l^{\gamma}$ (backward traveling), except that the source is emitting in the backward direction. These coefficients can be determined using the transfer matrix or scattering matrix approach [24,26].

Inserting Eqs. (10) and (11) in Eq. (9), and performing the integration over the thickness of the source layer, leads to:

$$\bar{\bar{\mathbf{W}}}(\mathbf{r}_1,\mathbf{r}_2,\omega)=\frac{4\omega^3\mu_0^2\varepsilon_0\varepsilon''(\omega)}{\pi}\Theta(\omega,T)\int_0^{\infty}\frac{KdK}{4\pi^2}\int_0^{2\pi}d\phi\,e^{iK(\cos(\phi)\Delta x+\sin(\phi)\Delta y)}\tau(K,\phi,z_1,z_2,\omega) \quad (14)$$

The transmission function $\tau$ is written in terms of TE- and TM-polarized unit vectors as follows:

$$\tau(K,\phi,z_1,z_2,\omega) = \tau_1 \hat{\mathbf{s}} \otimes \hat{\mathbf{s}} + \tau_2 \hat{\mathbf{p}}_l^+ \otimes \hat{\mathbf{p}}_l^+ + \tau_3 \hat{\mathbf{p}}_l^- \otimes \hat{\mathbf{p}}_l^- + \tau_4 \hat{\mathbf{p}}_l^+ \otimes \hat{\mathbf{p}}_l^- + \tau_5 \hat{\mathbf{p}}_l^- \otimes \hat{\mathbf{p}}_l^+ \tag{15}$$

where the $\tau_i$ components are independent of $\phi$ and are given by:

$$\tau_1(K,z_1,z_2,\omega) = \frac{1}{4|k_{zs}|^2} \left( \begin{array}{l} e^{ik_{zl}\Delta z_1} e^{-ik_{zl}^*\Delta z_2} \left( \begin{array}{l} \dfrac{1}{2k_{zs}''} \dfrac{e^{2t_s k_{zs}''}-1}{e^{2t_s k_{zs}''}} \left( |A_l^{TE}|^2 e^{2t_s k_{zs}''} + |C_l^{TE}|^2 \right) \\ -\dfrac{i}{2k_{zs}'} \dfrac{e^{2t_s ik_{zs}'}-1}{e^{2t_s ik_{zs}'}} \left( A_l^{TE} C_l^{TE*} + C_l^{TE} A_l^{TE*} e^{2t_s ik_{zs}'} \right) \end{array} \right) \\ + e^{-ik_{zl}\Delta z_1} e^{ik_{zl}^*\Delta z_2} \left( \begin{array}{l} \dfrac{1}{2k_{zs}''} \dfrac{e^{2t_s k_{zs}''}-1}{e^{2t_s k_{zs}''}} \left( |B_l^{TE}|^2 e^{2t_s k_{zs}''} + |D_l^{TE}|^2 \right) \\ -\dfrac{i}{2k_{zs}'} \dfrac{e^{2t_s ik_{zs}'}-1}{e^{2t_s ik_{zs}'}} \left( B_l^{TE} D_l^{TE*} + D_l^{TE} B_l^{TE*} e^{2t_s ik_{zs}'} \right) \end{array} \right) \\ + e^{ik_{zl}\Delta z_1} e^{ik_{zl}^*\Delta z_2} \left( \begin{array}{l} \dfrac{1}{2k_{zs}''} \dfrac{e^{2t_s k_{zs}''}-1}{e^{2t_s k_{zs}''}} \left( A_l^{TE} B_l^{TE*} e^{2t_s k_{zs}''} + C_l^{TE} D_l^{TE*} \right) \\ -\dfrac{i}{2k_{zs}'} \dfrac{e^{2t_s ik_{zs}'}-1}{e^{2t_s ik_{zs}'}} \left( A_l^{TE} D_l^{TE*} + C_l^{TE} B_l^{TE*} e^{2t_s ik_{zs}'} \right) \end{array} \right) \\ + e^{-ik_{zl}\Delta z_1} e^{-ik_{zl}^*\Delta z_2} \left( \begin{array}{l} \dfrac{1}{2k_{zs}''} \dfrac{e^{2t_s k_{zs}''}-1}{e^{2t_s k_{zs}''}} \left( B_l^{TE} A_l^{TE*} e^{2t_s k_{zs}''} + D_l^{TE} C_l^{TE*} \right) \\ -\dfrac{i}{2k_{zs}'} \dfrac{e^{2t_s ik_{zs}'}-1}{e^{2t_s ik_{zs}'}} \left( B_l^{TE} C_l^{TE*} + D_l^{TE} A_l^{TE*} e^{2t_s ik_{zs}'} \right) \end{array} \right) \end{array} \right) \tag{16}$$

$$\tau_2(K,z_1,z_2,\omega) = \frac{e^{ik_{zl}\Delta z_1} e^{-ik_{zl}^*\Delta z_2}}{4|k_{zs}|^2} \left( \begin{array}{l} \dfrac{1}{2k_{zs}''} \dfrac{e^{2t_s k_{zs}''}-1}{e^{2t_s k_{zs}''}} \dfrac{K^2+|k_{zs}|^2}{|k_s|^2} \left( |A_l^{TM}|^2 e^{2t_s k_{zs}''} + |C_l^{TM}|^2 \right) \\ -\dfrac{i}{2k_{zs}'} \dfrac{e^{2t_s ik_{zs}'}-1}{e^{2t_s ik_{zs}'}} \dfrac{K^2-|k_{zs}|^2}{|k_s|^2} \left( A_l^{TM} C_l^{TM*} + A_l^{TM*} C_l^{TM} e^{2t_s ik_{zs}'} \right) \end{array} \right) \tag{17}$$

$$\tau_3(K,z_1,z_2,\omega) = \frac{e^{-ik_{zl}\Delta z_1}e^{ik_{zl}^*\Delta z_2}}{4|k_{zs}|^2}\left(\begin{array}{l}\dfrac{1}{2k_{zs}''}\dfrac{e^{2t_s k_{zs}''}-1}{e^{2t_s k_{zs}''}}\dfrac{K^2+|k_{zs}|^2}{|k_s|^2}\left(|B_l^{TM}|^2 e^{2t_s k_{zs}''}+|D_l^{TM}|^2\right)\\ -\dfrac{i}{2k_{zs}'}\dfrac{e^{2t_s ik_{zs}'}-1}{e^{2t_s ik_{zs}'}}\dfrac{K^2-|k_{zs}|^2}{|k_s|^2}\left(B_l^{TM}D_l^{TM*}+B_l^{TM*}D_l^{TM}e^{2t_s ik_{zs}'}\right)\end{array}\right) \quad (18)$$

$$\tau_4(K,z_1,z_2,\omega) = \frac{e^{ik_{zl}\Delta z_1}e^{ik_{zl}^*\Delta z_2}}{4|k_{zs}|^2}\left(\begin{array}{l}\dfrac{1}{2k_{zs}''}\dfrac{e^{2t_s k_{zs}''}-1}{e^{2t_s k_{zs}''}}\dfrac{K^2+|k_{zs}|^2}{|k_s|^2}\left(A_l^{TM}B_l^{TM*}e^{2t_s k_{zs}''}+C_l^{TM}D_l^{TM*}\right)\\ -\dfrac{i}{2k_{zs}'}\dfrac{e^{2t_s ik_{zs}'}-1}{e^{2t_s ik_{zs}'}}\dfrac{K^2-|k_{zs}|^2}{|k_s|^2}\left(A_l^{TM}D_l^{TM*}+B_l^{TM*}C_l^{TM}e^{2t_s ik_{zs}'}\right)\end{array}\right) \quad (19)$$

$$\tau_5(K,z_1,z_2,\omega) = \frac{e^{-ik_{zl}\Delta z_1}e^{-ik_{zl}^*\Delta z_2}}{4|k_{zs}|^2}\left(\begin{array}{l}\dfrac{1}{2k_{zs}''}\dfrac{e^{2t_s k_{zs}''}-1}{e^{2t_s k_{zs}''}}\dfrac{K^2+|k_{zs}|^2}{|k_s|^2}\left(A_l^{TM*}B_l^{TM}e^{2t_s k_{zs}''}+C_l^{TM*}D_l^{TM}\right)\\ -\dfrac{i}{2k_{zs}'}\dfrac{e^{2t_s ik_{zs}'}-1}{e^{2t_s ik_{zs}'}}\dfrac{K^2-|k_{zs}|^2}{|k_s|^2}\left(B_l^{TM}C_l^{TM*}+A_l^{TM*}D_l^{TM}e^{2t_s ik_{zs}'}\right)\end{array}\right) \quad (20)$$

The outer products of TE- and TM-polarized unit vectors in Eq. (15) are explicitly expressed in terms of the azimuthal angle, $\phi$, as follows:

$$\hat{\mathbf{s}}\otimes\hat{\mathbf{s}} = \begin{pmatrix} \sin^2(\phi) & -\sin(\phi)\cos(\phi) & 0 \\ -\sin(\phi)\cos(\phi) & \cos^2(\phi) & 0 \\ 0 & 0 & 0 \end{pmatrix} \quad (21)$$

$$\hat{\mathbf{p}}_l^+\otimes\hat{\mathbf{p}}_l^+ = \frac{1}{|k_l|^2}\begin{pmatrix} |k_{zl}|^2\cos^2(\phi) & |k_{zl}|^2\sin(\phi)\cos(\phi) & -k_{zl}K\cos(\phi) \\ |k_{zl}|^2\sin(\phi)\cos(\phi) & |k_{zl}|^2\sin^2(\phi) & -k_{zl}K\sin(\phi) \\ -k_{zl}^*K\cos(\phi) & -k_{zl}^*K\sin(\phi) & K^2 \end{pmatrix} \quad (22)$$

$$\hat{\mathbf{p}}_l^- \otimes \hat{\mathbf{p}}_l^- = \frac{1}{|k_l|^2} \begin{pmatrix} |k_{zl}|^2 \cos^2(\phi) & |k_{zl}|^2 \sin(\phi)\cos(\phi) & k_{zl} K \cos(\phi) \\ |k_{zl}|^2 \sin(\phi)\cos(\phi) & |k_{zl}|^2 \sin^2(\phi) & k_{zl} K \sin(\phi) \\ k_{zl}^* K \cos(\phi) & k_{zl}^* K \sin(\phi) & K^2 \end{pmatrix} \quad (23)$$

$$\hat{\mathbf{p}}_l^+ \otimes \hat{\mathbf{p}}_l^- = \frac{1}{|k_l|^2} \begin{pmatrix} -|k_{zl}|^2 \cos^2(\phi) & -|k_{zl}|^2 \sin(\phi)\cos(\phi) & -k_{zl} K \cos(\phi) \\ -|k_{zl}|^2 \sin(\phi)\cos(\phi) & -|k_{zl}|^2 \sin^2(\phi) & -k_{zl} K \sin(\phi) \\ k_{zl}^* K \cos(\phi) & k_{zl}^* K \sin(\phi) & K^2 \end{pmatrix} \quad (24)$$

$$\hat{\mathbf{p}}_l^- \otimes \hat{\mathbf{p}}_l^- = \frac{1}{|k_l|^2} \begin{pmatrix} -|k_{zl}|^2 \cos^2(\phi) & -|k_{zl}|^2 \sin(\phi)\cos(\phi) & k_{zl} K \cos(\phi) \\ -|k_{zl}|^2 \sin(\phi)\cos(\phi) & -|k_{zl}|^2 \sin^2(\phi) & k_{zl} K \sin(\phi) \\ -k_{zl}^* K \cos(\phi) & -k_{zl}^* K \sin(\phi) & K^2 \end{pmatrix} \quad (25)$$

Inspection of Eqs. (21) to (25) reveals that six different integrations over the azimuthal angle need to be performed in the cross-spectral density tensor (Eq. (14)). These integrations are carried out analytically for the first time in this paper, and the resulting expressions are:

$$I_1 = \int_0^{2\pi} d\phi\, e^{iK(\cos(\phi)\Delta x + \sin(\phi)\Delta y)} = 2\pi J_0\left(K\sqrt{\Delta x^2 + \Delta y^2}\right) \quad (26)$$

$$I_2 = \int_0^{2\pi} d\phi\, \sin^2(\phi) e^{iK(\cos(\phi)\Delta x + \sin(\phi)\Delta y)} = -\frac{4\Delta y^2 \pi J_1\left(K\sqrt{\Delta x^2 + \Delta y^2}\right)}{K\sqrt{(\Delta x^2 + \Delta y^2)^3}} + \frac{2\pi J_1\left(K\sqrt{\Delta x^2 + \Delta y^2}\right)}{K\sqrt{\Delta x^2 + \Delta y^2}} + \frac{2\Delta y^2 \pi J_0\left(K\sqrt{\Delta x^2 + \Delta y^2}\right)}{\Delta x^2 + \Delta y^2} \quad (27)$$

$$I_3 = \int_0^{2\pi} d\phi \sin(\phi)\cos(\phi) e^{iK(\cos(\phi)\Delta x + \sin(\phi)\Delta y)} = -\frac{4\Delta x \Delta y \pi J_1\left(K\sqrt{\Delta x^2 + \Delta y^2}\right)}{K\sqrt{(\Delta x^2 + \Delta y^2)^3}}$$
$$+ \frac{2\Delta x \Delta y \pi J_0\left(K\sqrt{\Delta x^2 + \Delta y^2}\right)}{\Delta x^2 + \Delta y^2} \quad (28)$$

$$I_4 = \int_0^{2\pi} d\phi \cos^2(\phi) e^{iK(\cos(\phi)\Delta x + \sin(\phi)\Delta y)} = -\frac{4\Delta x^2 \pi J_1\left(K\sqrt{\Delta x^2 + \Delta y^2}\right)}{K\sqrt{(\Delta x^2 + \Delta y^2)^3}} + \frac{2\pi J_1\left(K\sqrt{\Delta x^2 + \Delta y^2}\right)}{K\sqrt{\Delta x^2 + \Delta y^2}}$$
$$+ \frac{2\Delta x^2 \pi J_0\left(K\sqrt{\Delta x^2 + \Delta y^2}\right)}{\Delta x^2 + \Delta y^2} \quad (29)$$

$$I_5 = \int_0^{2\pi} d\phi \cos(\phi) e^{iK(\cos(\phi)\Delta x + \sin(\phi)\Delta y)} = \frac{2i\Delta x \pi J_1\left(K\sqrt{\Delta x^2 + \Delta y^2}\right)}{\sqrt{\Delta x^2 + \Delta y^2}} \quad (30)$$

$$I_6 = \int_0^{2\pi} d\phi \sin(\phi) e^{iK(\cos(\phi)\Delta x + \sin(\phi)\Delta y)} = \frac{2i\Delta y \pi J_1\left(K\sqrt{\Delta x^2 + \Delta y^2}\right)}{\sqrt{\Delta x^2 + \Delta y^2}} \quad (31)$$

where the Bessel functions of the first kind (zeroth and first order) are denoted by $J_0$ and $J_1$. The importance of performing analytically the integrations over the azimuthal angle is twofold. First, it substantially accelerates numerical calculation of the cross-spectral density tensor. Second, it avoids potential inaccuracy in the cross-spectral density tensor, as the numerical integration over the azimuthal angle is highly sensitive to the $\phi$-discretization.

Integrations of the outer products of TE- and TM-polarized unit vectors (Eqs. (21) to (25)) over $\phi$ in the cross-spectral density tensor (Eq. (14)) are then carried out using Eqs. (26) to (31):

$$\Phi_1(K,\Delta x,\Delta y) = \int_0^{2\pi} d\phi\, \hat{\mathbf{s}} \otimes \hat{\mathbf{s}}\, e^{iK(\cos(\phi)\Delta x+\sin(\phi)\Delta y)} = \begin{pmatrix} I_2 & -I_3 & 0 \\ -I_3 & I_4 & 0 \\ 0 & 0 & 0 \end{pmatrix} \quad (32)$$

$$\Phi_2(K,\Delta x,\Delta y,\omega) = \int_0^{2\pi} d\phi\, \hat{\mathbf{p}}_l^+ \otimes \hat{\mathbf{p}}_l^+\, e^{iK(\cos(\phi)\Delta x+\sin(\phi)\Delta y)} = \frac{1}{|k_l|^2}\begin{pmatrix} |k_{zl}|^2 I_4 & |k_{zl}|^2 I_3 & -k_{zl} K I_5 \\ |k_{zl}|^2 I_3 & |k_{zl}|^2 I_2 & -k_{zl} K I_6 \\ -k_{zl}^* K I_5 & -k_{zl}^* K I_6 & K^2 I_1 \end{pmatrix} \quad (33)$$

$$\Phi_3(K,\Delta x,\Delta y,\omega) = \int_0^{2\pi} d\phi\, \hat{\mathbf{p}}_l^- \otimes \hat{\mathbf{p}}_l^-\, e^{iK(\cos(\phi)\Delta x+\sin(\phi)\Delta y)} = \frac{1}{|k_l|^2}\begin{pmatrix} |k_{zl}|^2 I_4 & |k_{zl}|^2 I_3 & k_{zl} K I_5 \\ |k_{zl}|^2 I_3 & |k_{zl}|^2 I_2 & k_{zl} K I_6 \\ k_{zl}^* K I_5 & k_{zl}^* K I_6 & K^2 I_1 \end{pmatrix} \quad (34)$$

$$\Phi_4(K,\Delta x,\Delta y,\omega) = \int_0^{2\pi} d\phi\, \hat{\mathbf{p}}_l^+ \otimes \hat{\mathbf{p}}_l^-\, e^{iK(\cos(\phi)\Delta x+\sin(\phi)\Delta y)} = \frac{1}{|k_l|^2}\begin{pmatrix} -|k_{zl}|^2 I_4 & -|k_{zl}|^2 I_3 & -k_{zl} K I_5 \\ -|k_{zl}|^2 I_3 & -|k_{zl}|^2 I_2 & -k_{zl} K I_6 \\ k_{zl}^* K I_5 & k_{zl}^* K I_6 & K^2 I_1 \end{pmatrix} \quad (35)$$

$$\Phi_5(K,\Delta x,\Delta y,\omega) = \int_0^{2\pi} d\phi\, \hat{\mathbf{p}}_l^- \otimes \hat{\mathbf{p}}_l^+\, e^{iK(\cos(\phi)\Delta x+\sin(\phi)\Delta y)} = \frac{1}{|k_l|^2}\begin{pmatrix} -|k_{zl}|^2 I_4 & -|k_{zl}|^2 I_3 & k_{zl} K I_5 \\ -|k_{zl}|^2 I_3 & -|k_{zl}|^2 I_2 & k_{zl} K I_6 \\ -k_{zl}^* K I_5 & -k_{zl}^* K I_6 & K^2 I_1 \end{pmatrix} \quad (36)$$

The final expression of the cross-spectral density tensor is formulated as follows:

$$\bar{\bar{\mathbf{W}}}(\mathbf{r}_1,\mathbf{r}_2,\omega) = \frac{\omega^3 \mu_0^2 \varepsilon_0 \varepsilon''(\omega)}{\pi^3}\Theta(\omega,T)\int_0^\infty K\, dK\left[\tau_1\Phi_1 + \tau_2\Phi_2 + \tau_3\Phi_3 + \tau_4\Phi_4 + \tau_5\Phi_5\right] \quad (37)$$

where $\tau_i$ and $\Phi_i$ are respectively given by Eqs. (16)-(20) and Eqs. (32)-(36). Eq. (37) is the main result of this paper. It can be applied to problems involving an arbitrary number of layers, and it only contains one integration over the parallel wavevector $K$. Note that explicit expressions for the nine components of the cross-spectral density tensor are provided in the Appendix.

## III. RESULTS

The general formalism presented in Section II is applied hereafter for the special case shown in Fig. 3, where a film heat source emitting in free space is coated on a semi-infinite substrate. The film is made of SiC supporting SPhPs in the infrared, while the substrate is assumed to be non-emitting. The goal is to analyze the impacts of SPhP coupling within the SiC film and the optical properties of the substrate on the spatial correlation of the electric field in free space. The spatial correlation of the electric field is calculated in free space between two locations, $\mathbf{r}_1 = (x_1, y, z)$ and $\mathbf{r}_2 = (x_2, y, z)$, located in the $x$-$z$ plane ($\Delta y = 0$) and at a distance $\Delta z$ ($= \Delta z_1 = \Delta z_2$) of $0.05\lambda$ with respect to the interface 1-2.

### A. Cross-spectral density tensor for an emitting film coated on a non-emitting substrate

The source layer is the film ($s = 1$) and the cross-spectral density tensor is calculated in free space ($l = 2$). The semi-infinite substrate onto which the film is coated is labeled as medium 0. The field amplitude coefficients in free space needed to calculate the transmission functions (Eqs. (16) to (20)) are given by [26]:

$$A_2^\gamma = \frac{t_{12}^\gamma}{1 + r_{01}^\gamma r_{12}^\gamma e^{2it_1 k_{z1}}} e^{it_1 k_{z1}} \tag{38}$$

$$B_2^\gamma = 0 \tag{39}$$

$$C_2^\gamma = -r_{01}^\gamma A_2^\gamma \tag{40}$$

$$D_2^\gamma = 0 \tag{41}$$

where the TE- and TM-polarized Fresnel reflection and transmission coefficients at the interface separating media $i$ and $j$ are given by $r_{ij}^{TE} = (k_{zi} - k_{zj})/(k_{zi} + k_{zj})$, $r_{ij}^{TM} = (\varepsilon_j k_{zi} - \varepsilon_i k_{zj})/(\varepsilon_j k_{zi} + \varepsilon_i k_{zj})$, $t_{ij}^{TE} = 2k_{zi}/(k_{zi} + k_{zj})$, and $t_{ij}^{TM} = \sqrt{\varepsilon_i/\varepsilon_j}(2\varepsilon_j k_{zi})/(\varepsilon_j k_{zi} + \varepsilon_i k_{zj})$ [27]. Substitution of the field amplitude coefficients into Eqs. (16) to (20) leads to the following non-zero transmission functions:

$$\tau_1 = \frac{e^{i(k_{z2}\Delta z_1 - k_{z2}^*\Delta z_2)}}{4|k_{z1}|^2} \frac{|t_{12}^{TE}|^2}{|1 + r_{01}^{TE} r_{12}^{TE} e^{2it_1 k_{z1}}|^2} \times \\
\left( \frac{1}{2k_{z1}''} \frac{e^{2t_1 k_{z1}''} - 1}{e^{2t_1 k_{z1}''}} \left(1 + |r_{01}^{TE}|^2 e^{-2t_1 k_{z1}''}\right) + \frac{i}{2k_{z1}'} \frac{e^{2it_1 k_{z1}'} - 1}{e^{2it_1 k_{z1}'}} \left( r_{01}^{TE} e^{2it_1 k_{z1}'} + r_{01}^{TE*} \right) e^{-2t_1 k_{z1}''} \right) \tag{42}$$

$$\tau_2 = \frac{e^{i(k_{z2}\Delta z_1 - k_{z2}^*\Delta z_2)}}{4|k_{z1}|^2} \frac{|t_{12}^{TM}|^2}{|1 + r_{01}^{TM} r_{12}^{TM} e^{2it_1 k_{z1}}|^2} \times \\
\left( \frac{1}{2k_{z1}''} \frac{e^{2t_1 k_{z1}''} - 1}{e^{2t_1 k_{z1}''}} \frac{K^2 + |k_{z1}|^2}{|k_1|^2} \left(1 + |r_{01}^{TM}|^2 e^{-2t_1 k_{z1}''}\right) \right. \\
\left. + \frac{i}{2k_{z1}'} \frac{e^{2it_1 k_{z1}'} - 1}{e^{2it_1 k_{z1}'}} \frac{K^2 - |k_{z1}|^2}{|k_1|^2} \left( r_{01}^{TM} e^{2t_1 ik_{z1}'} + r_{01}^{TM*} \right) e^{-2t_1 k_{z1}''} \right) \tag{43}$$

Here, since the spatial correlation of the electric field is calculated in the *x-z* plane such that $\Delta y = 0$, $I_3$ and $I_6$ are zero (see Eqs. (28) and (31)) resulting in five non-zero components of the cross-spectral density tensor:

$$W_{xx}(\mathbf{r}_1,\mathbf{r}_2,\omega) = \frac{\omega^3 \mu_0^2 \varepsilon_0 \varepsilon''(\omega)}{\pi^3} \Theta(\omega,T) \int_0^\infty K\, dK \left[ I_2 \tau_1 + \frac{|k_{z2}|^2}{|k_2|^2} I_4 \tau_2 \right] \quad (44)$$

$$W_{xz}(\mathbf{r}_1,\mathbf{r}_2,\omega) = \frac{\omega^3 \mu_0^2 \varepsilon_0 \varepsilon''(\omega)}{\pi^3} \Theta(\omega,T) \int_0^\infty K\, dK \left[ -\frac{k_{z2} K}{|k_2|^2} I_5 \tau_2 \right] \quad (45)$$

$$W_{yy}(\mathbf{r}_1,\mathbf{r}_2,\omega) = \frac{\omega^3 \mu_0^2 \varepsilon_0 \varepsilon''(\omega)}{\pi^3} \Theta(\omega,T) \int_0^\infty K\, dK \left[ I_4 \tau_1 + \frac{|k_{z2}|^2}{|k_2|^2} I_2 \tau_2 \right] \quad (46)$$

$$W_{zx}(\mathbf{r}_1,\mathbf{r}_2,\omega) = \frac{\omega^3 \mu_0^2 \varepsilon_0 \varepsilon''(\omega)}{\pi^3} \Theta(\omega,T) \int_0^\infty K\, dK \left[ -\frac{k_{z2}^* K}{|k_2|^2} I_5 \tau_2 \right] \quad (47)$$

$$W_{zz}(\mathbf{r}_1,\mathbf{r}_2,\omega) = \frac{\omega^3 \mu_0^2 \varepsilon_0 \varepsilon''(\omega)}{\pi^3} \Theta(\omega,T) \int_0^\infty K\, dK \left[ \frac{K^2}{|k_2|^2} I_1 \tau_2 \right] \quad (48)$$

Note that the term $e^{2t_s k''_{zs}}$ in the transmission coefficients (Eqs. (16) to (20)) leads to numerical instability when the source is thick and/or when the parallel wavevector $K$ is large. Here, the exponential term is eliminated by substituting $\left|A_2^\gamma\right|^2 = \dfrac{\left|t_{12}^\gamma\right|^2}{\left|1 + r_{01}^\gamma r_{12}^\gamma e^{2it_1 k_{z1}}\right|^2} e^{-2t_1 k''_{z1}}$ in Eqs. (16) and (17). In that way, the resulting transmission functions given by Eqs. (42) and (43) do not contain the term $e^{2t_1 k''_{z1}}$.

In the limiting case of a bulk heat source ($t_1 \to \infty$), the transmission functions $\tau_1$ and $\tau_2$ are simplified as follows using $\lim\limits_{t_1 \to \infty} \frac{e^{2t_1 k''_{z1}} - 1}{e^{2t_1 k''_{z1}}} \to 1$ and $\lim\limits_{t_1 \to \infty} e^{2it_1 k_{z1}} \to 0$:

$$\tau_1 = \frac{e^{i(k_{z2}\Delta z_1 - k^*_{z2}\Delta z_2)}}{8k''_{z1}|k_{z1}|^2}|t^{TE}_{12}|^2 \tag{49}$$

$$\tau_2 = \frac{e^{i(k_{z2}\Delta z_1 - k^*_{z2}\Delta z_2)}}{8k''_{z1}|k_{z1}|^2} \frac{K^2 + |k_{z1}|^2}{|k_1|^2}|t^{TM}_{12}|^2 \tag{50}$$

The same result for a bulk heat source can be obtained by using the field amplitude coefficients $A^\gamma_2 = t^\gamma_{12}$, and $B^\gamma_2 = C^\gamma_2 = D^\gamma_2 = 0$, derived via the transfer matrix method for a single interface, and by performing the integration over the $dz'$ in Eq. (9) from $-\infty$ to 0.

### B. Analysis of cross-spectral density tensor

The cross-spectral density tensor is calculated at a single wavelength $\lambda$ of 11.36 μm (frequency $\omega$ of $1.658 \times 10^{14}$ rad/s) located within the Reststrahlen band where SiC supports SPhPs. At this wavelength, the dielectric function of SiC is $\varepsilon_1 = -7.514 + 0.408i$ [28]. This wavelength has been selected for the purpose of verification with Ref. [8] where the spatial correlation of the thermal electromagnetic near field of a SiC bulk was reported. Figure 4(a) shows the element $W_{xx}$ of the cross-spectral density tensor for 10-nm and 200-nm-thick films surrounded by vacuum, and for a bulk SiC thermal source, as a function of the length $|\mathbf{r}_1 - \mathbf{r}_2|$. Propagating and evanescent modes for both TM and TE polarizations are considered when calculating the cross-spectral density tensor. The other elements of the cross-spectral density tensor, namely $W_{yy}$, $W_{zz}$, $W_{xz}$ and $W_{zx}$, are provided in Fig. S1 of the Supplemental Material [29].

While it is known that a bulk heat source made of SiC results in a large spatial correlation length at a wavelength of 11.36 µm (~40$\lambda$) owing to SPhPs, Fig. 4(a) shows that the spatial correlation length of a SiC heat source decreases substantially as its thickness decreases. Despite the presence of SPhPs, the spatial correlation length for the 10-nm-thick film is comparable to that of a blackbody (~0.5$\lambda$) [8,18]. This behavior can be explained by analyzing the TM-polarized transmission coefficient, $\tau_2$, plotted in Fig. 4(b) as a function of the normalized parallel wavevector, $\overline{K}$ (= $K/k_0$), and the heat source thickness, $t_1$. Note that only TM polarization is analyzed here, since SPhPs can only be excited in that polarization state [2]. Resonance of the transmission function occurs near a single narrow $K$-band for film thicknesses approximately equal to or larger than 2 µm. For thinner films, resonance of the transmission function splits into two branches, a narrow one near a $\overline{K}$ value of 1, and a second one spreading out over a wide $K$-band. This is due to coupling of SPhPs, existing at both SiC-vacuum interfaces, within the film leading to a resonance splitting into anti-symmetric, $\omega^+$, and symmetric, $\omega^-$, modes [30]. SPhP dispersion relation is determined in the limit that $\tau_2 \to \infty$, which is achieved when $1 + r_{01}^{TM} r_{12}^{TM} e^{2it_1 k_{z1}} = 0$ (see Eq. (43)). For the specific case of a film surrounded by vacuum, the dispersion relations of anti-symmetric and symmetric modes respectively satisfy the following equations [31]:

$$\tanh\left(\frac{k_{z1} t_1}{2}\right) = -\frac{k_{z0} \varepsilon_1}{k_{z1}} \tag{51}$$

$$\tanh\left(\frac{k_{z1} t_1}{2}\right) = -\frac{k_{z1}}{k_{z0} \varepsilon_1} \tag{52}$$

SPhP dispersion relations, and thus resonance splitting, depend on the film thickness, $t_1$, and the dielectric function, $\varepsilon_0$ and $\varepsilon_2$ (= 1 for vacuum), of the media surrounding the film in addition to the dielectric function of SiC, $\varepsilon_1$. Using the definition of penetration depth, $\delta_j \approx |k_{zj}|^{-1}$ and $k_{zj} \approx iK$ in the electrostatic limit, the maximum contributing parallel wavevector to near-field thermal emission at $\Delta z$ (= 0.05$\lambda$) is estimated as $K_{max} \approx (\Delta z)^{-1}$ ($\overline{K}_{max} \approx [k_0(0.05\lambda)]^{-1} = 10/\pi$). The largest contributing parallel wavevector is identified as a vertical dotted line in Fig. 4(b). Larger parallel wavevectors cannot contribute to near-field thermal emission at $\Delta z$ since their penetration depths are smaller than $\Delta z$. Due to the fact that $K_{max}$ dominates near-field thermal emission [30], resonance splitting of the transmission function is not visible when the film is much thicker than $\Delta z$ (568 nm), which is in good agreement with Fig. 4(b). When $t_1 \gg \Delta z$, SPhP dispersion relation reduces to [30]:

$$K_{SPhP} = \sqrt{\frac{\varepsilon_1}{\varepsilon_1 + 1}} \qquad (53)$$

which is the dispersion relation of a SiC-vacuum interface.

For the 10-nm-thick film, SPhP coupling within the heat source is strong since $t_1 \ll \Delta z$. This strong coupling pushes the symmetric mode to a $\overline{K}$ value of ~40, which largely exceeds $\overline{K}_{max} \approx 10/\pi$, meaning that the penetration depth of the symmetric mode is too small to contribute to near-field thermal emission at 0.05$\lambda$. In addition, the anti-symmetric resonance is very weak and does not dominate near-field thermal emission. These combined effects lead to a spatially incoherent heat source. By analogy with temporally coherent heat sources emitting in a narrow frequency $\omega$-band, spatial coherence is characterized by large thermal emission in a narrow wavevector $K$-band. For the 200-nm-thick film, the spreading of the symmetric mode over a large $K$-band combined with the weak resonance of the anti-symmetric mode result in a spatial correlation length of ~4$\lambda$. This is significantly shorter than the spatial correlation length of a SiC bulk, where a single SPhP resonance dominating near-field thermal emission occurs within a narrow *K*-band.

A bulk SiC heat source does not exhibit spatial coherence when $\varepsilon_1' = -1$ at a wavelength of 10.55 µm, which corresponds to resonance of a single SiC-vacuum interface [30]. The element $W_{xx}$ of the cross-spectral density tensor for this case is shown in Fig. S2 of the Supplemental Material [29]. Near-field thermal emission by a SiC bulk heat source is quasi-monochromatic around a wavelength of 10.55 µm, which is a manifestation of a high degree of temporal coherence. Yet, at this wavelength, a large number of *K*-modes are excited, thus resulting in thermal emission that is spatially incoherent.

Carminati and Greffet [8] concluded that surface polaritons correlate the thermal near field over distances on the order of their propagation length. For a SiC bulk, SPhP propagation length is estimated at $36\lambda$ using $1/\text{Im}(K_{SPhP})$ [20], where $K_{SPhP}$ satisfies Eq. (53). SPhP propagation length is indeed similar to the spatial correlation length of $\sim 40\lambda$. For thin films supporting SPhPs like SiC, the anti-symmetric mode results in a propagation length far exceeding that of its bulk counterpart (the symmetric mode results in a propagation length smaller than that of a bulk) [31]. This effect has been exploited for increasing the in-plane thermal conductivity of thin films supporting SPhPs in the infrared [32-34]. For a 10-nm-thick SiC film, the propagation length of the anti-symmetric mode, estimated via $K_{SPhP}$ satisfying Eq. (51), is $1/\text{Im}(K_{SPhP}) \approx 2.5 \times 10^6 \lambda$ [35]. Yet, this extremely large propagation length does not translate into a large spatial correlation length, as shown in Fig. 4(a). Again, this is due to the fact that near-field thermal emission by a 10-nm-thick SiC film at a distance $0.05\lambda$ and a wavelength of 11.36 µm is not resonant around the anti-symmetric mode. As such, it is safer to conclude that for materials supporting surface polaritons, a large spatial correlation length implies a long surface polariton propagation length. However, a long surface polariton propagation length is a necessary but insufficient condition for obtaining a large spatial correlation length.

Finally, it is shown hereafter that the spatial coherence of a thin SiC heat source can be modulated via a substrate. Figure 5 displays the element $W_{xx}$ of the cross-spectral density tensor as a function of the length $|\mathbf{r}_1 - \mathbf{r}_2|$ at a wavelength of 11.36 μm for a 10-nm-thick SiC film coated on dielectric ($\varepsilon_0 = 2 + 0.001i$) and metallic ($\varepsilon_0 = -2 + 0.1i$) substrates assumed to be non-emitting. For the dielectric substrate, SPhP coupling within the SiC film occurs thus leading to a small spatial correlation length, as explained previously. For a metallic substrate, the spatial correlation length is substantially increased to a value of ~ 12$\lambda$. This is explained by the fact that the real part of the dielectric functions of both the substrate and the film are negative, thus preventing the generation of surface polaritons at the interface 0-1 [31]. As such, only the interface 1-2 delimiting SiC and vacuum supports SPhPs. The spatial coherence of the SiC film coated on a metallic substrate is similar to that of a bulk SiC heat source. The smaller spatial correlation length for the film is explained by the presence of the metallic substrate that decreases SPhP propagation length owing to losses. If the losses of the metallic substrate are further increased, SPhP propagation length and thus the spatial correlation length decrease. For instance, for a 10-nm-thick film on a metallic substrate described by a dielectric function $\varepsilon_0 = -2 + 10i$, the spatial correlation length is only ~0.2$\lambda$.

The results of section III are interesting from a practical standpoint. Indeed, they suggest that spatial coherence of a heat source can be modulated by allowing (or not) SPhP coupling at the substrate-film interface. A phase change material such as vanadium dioxide (VO$_2$) [36] could be used as a substrate for active control of thermal emission by allowing SPhP coupling when $VO_2$ is in the dielectric phase (incoherent emission), and by preventing SPhP coupling when $VO_2$ is in the metallic phase (coherent emission). In addition, the results of Fig. 5 demonstrate that spatial coherence of thin heat sources supporting surface polaritons is a strong function of the surrounding materials. As such, by coupling evanescent waves to the far field via, for instance, a grating, thermal emission by thin heat sources supporting surface polaritons could be exploited for sensing applications that would have the advantage of not requiring external illumination.

## IV. CONCLUSIONS

A general formulation for the cross-spectral density tensor enabling calculation of the spatial correlation of the thermally generated electromagnetic field in layered media has been derived. The formulation is based on fluctuational electrodynamics, such that it is applicable in both the far and near field of heat sources. The resulting cross-spectral density tensor is written in terms of a single integration over the parallel wavevector, as the angular integrations have been performed analytically. This results in higher accuracy for the calculation of the cross-spectral density tensor since the angular integrations are numerically unstable, in addition to reduced computational time. This is important not only for evaluating the spatial correlation length of heat sources, but also for numerical approaches such as the thermal discrete dipole approximation where the spatial correlation of the electric field needs to be calculated [37,38].

The formulation has been applied to calculate the spatial correlation length in the near field of a film heat source made of SiC at a wavelength of 11.36 μm, where SiC supports SPhPs. When suspended in vacuum, the spatial correlation length of the film heat source decreases as its thickness decreases due to SPhP coupling. For a 10-nm-thick film, the spatial correlation length is only ~0.5$\lambda$, which is similar to that of a blackbody, as opposed to ~40$\lambda$ for a bulk SiC heat source. Despite a long propagation length, the anti-symmetric mode resulting from SPhP coupling does not induce a large spatial correlation length due to its weak contribution to near-field thermal emission. It is, however, possible to control the spatial coherence of a thin SiC heat source via dielectric and metallic substrates respectively allowing and preventing SPhP coupling. This suggests that it may be possible to actively modulate thermal emission by using a phase change material substrate such as $VO_2$.

## ACKNOWLEDGMENTS

This work has been supported by the National Science Foundation (grant no. CBET-1253577). V.H. acknowledges the financial support from the Graduate Research Fellowship at the University of Utah.

# APPENDIX: COMPONENTS OF THE CROSS-SPECTRAL DENSITY TENSOR IN LAYERED MEDIA

The nine components of the cross-spectral density tensor (Eq. (37)) are:

$$W_{xx}(\mathbf{r}_1,\mathbf{r}_2,\omega) = \frac{\omega^3 \mu_0^2 \varepsilon_0 \varepsilon''(\omega)}{\pi^3} \Theta(\omega,T) \int_0^\infty K\, dK \left[ I_2 \tau_1 + \frac{|k_{zl}|^2}{|k_l|^2} I_4 (\tau_2 + \tau_3 - \tau_4 - \tau_5) \right] \tag{A1}$$

$$W_{xy}(\mathbf{r}_1,\mathbf{r}_2,\omega) = \frac{\omega^3 \mu_0^2 \varepsilon_0 \varepsilon''(\omega)}{\pi^3} \Theta(\omega,T) \int_0^\infty K\, dK \left[ -I_3 \tau_1 + \frac{|k_{zl}|^2}{|k_l|^2} I_3 (\tau_2 + \tau_3 - \tau_4 - \tau_5) \right] \tag{A2}$$

$$W_{xz}(\mathbf{r}_1,\mathbf{r}_2,\omega) = \frac{\omega^3 \mu_0^2 \varepsilon_0 \varepsilon''(\omega)}{\pi^3} \Theta(\omega,T) \int_0^\infty K\, dK \left[ \frac{k_{zl} K}{|k_l|^2} I_5 (-\tau_2 + \tau_3 - \tau_4 + \tau_5) \right] \tag{A3}$$

$$W_{yx}(\mathbf{r}_1,\mathbf{r}_2,\omega) = \frac{\omega^3 \mu_0^2 \varepsilon_0 \varepsilon''(\omega)}{\pi^3} \Theta(\omega,T) \int_0^\infty K\, dK \left[ -I_3 \tau_1 + \frac{|k_{zl}|^2}{|k_l|^2} I_3 (\tau_2 + \tau_3 - \tau_4 - \tau_5) \right] \tag{A4}$$

$$W_{yy}(\mathbf{r}_1,\mathbf{r}_2,\omega) = \frac{\omega^3 \mu_0^2 \varepsilon_0 \varepsilon''(\omega)}{\pi^3} \Theta(\omega,T) \int_0^\infty K\, dK \left[ I_4 \tau_1 + \frac{|k_{zl}|^2}{|k_l|^2} I_2 (\tau_2 + \tau_3 - \tau_4 - \tau_5) \right] \tag{A5}$$

$$W_{yz}(\mathbf{r}_1,\mathbf{r}_2,\omega) = \frac{\omega^3 \mu_0^2 \varepsilon_0 \varepsilon''(\omega)}{\pi^3} \Theta(\omega,T) \int_0^\infty K\, dK \left[ \frac{k_{zl} K}{|k_l|^2} I_6 (-\tau_2 + \tau_3 - \tau_4 + \tau_5) \right] \tag{A6}$$

$$W_{zx}(\mathbf{r}_1,\mathbf{r}_2,\omega) = \frac{\omega^3 \mu_0^2 \varepsilon_0 \varepsilon''(\omega)}{\pi^3} \Theta(\omega,T) \int_0^\infty K\, dK \left[ \frac{k_{zl}^* K}{|k_l|^2} I_5 (-\tau_2 + \tau_3 + \tau_4 - \tau_5) \right] \tag{A7}$$

$$W_{zy}(\mathbf{r}_1,\mathbf{r}_2,\omega) = \frac{\omega^3 \mu_0^2 \varepsilon_0 \varepsilon''(\omega)}{\pi^3} \Theta(\omega,T) \int_0^\infty K\, dK \left[ \frac{k_{zl}^* K}{|k_l|^2} I_6(-\tau_2+\tau_3+\tau_4-\tau_5) \right] \quad \text{(A8)}$$

$$W_{zz}(\mathbf{r}_1,\mathbf{r}_2,\omega) = \frac{\omega^3 \mu_0^2 \varepsilon_0 \varepsilon''(\omega)}{\pi^3} \Theta(\omega,T) \int_0^\infty K\, dK \left[ \frac{K^2}{|k_l|^2} I_1(\tau_2+\tau_3+\tau_4+\tau_5) \right] \quad \text{(A9)}$$

where $\tau_i$ and $I_i$ are respectively given by Eqs. (16)-(20) and Eqs. (26)-(31).

**FIGURES**

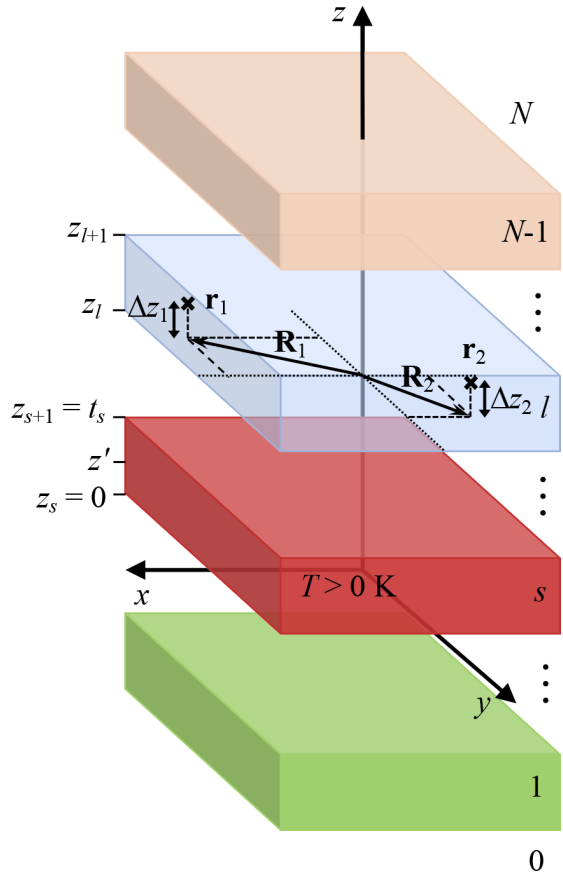

FIG. 1. Schematic of the geometry considered consisting of $N$-1 layers of finite thicknesses sandwiched between two semi-finite layers labeled media 0 and $N$. The spatial correlation of the electric field is calculated in layer $l$ between locations $\mathbf{r}_1$ and $\mathbf{r}_2$ due to thermal emission by layer $s$ maintained at temperature $T$.

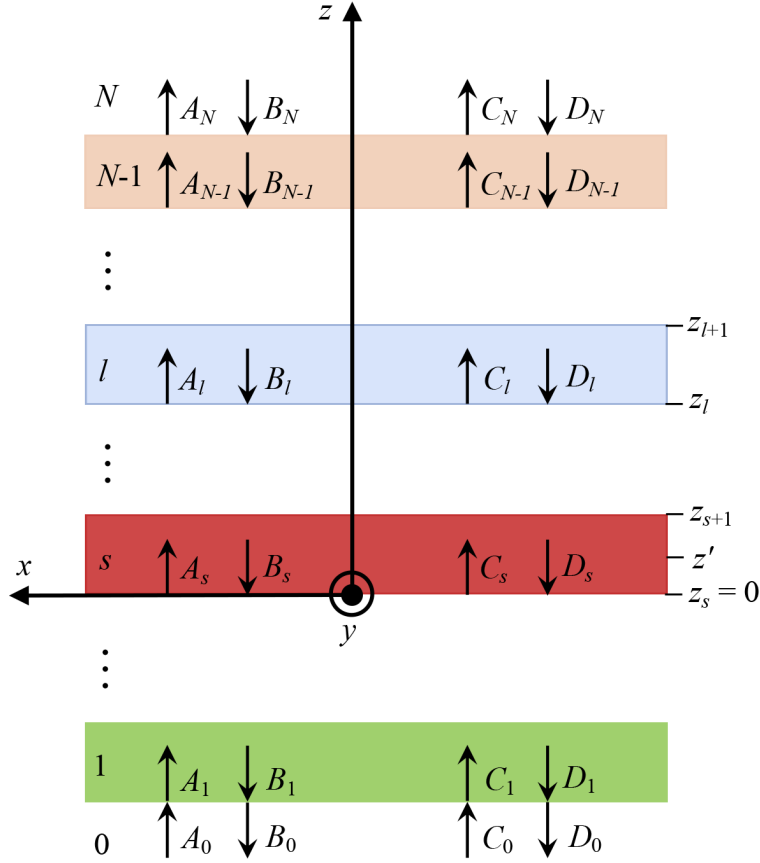

FIG. 2. Projection of the layered geometry shown in Fig. 1 in the *x-z* plane. The coefficients $A_i$ and $B_i$ are respectively the amplitudes of forward and backward traveling waves, with respect to the *z*-direction, in layer *i* generated by a source emitting in the forward direction. The coefficients $C_i$ and $D_i$ are respectively the amplitudes of forward and backward traveling waves, with respect to the *z*-direction, in layer *i* generated by a source emitting in the backward direction. The superscript $\gamma$ in the field amplitude coefficients representing the polarization state (TE or TM) is omitted in the figure for clarity.

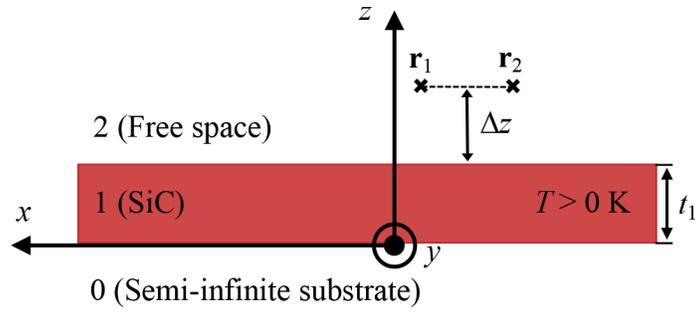

FIG. 3. Schematic of the emitting film coated on a non-emitting, semi-infinite substrate. The spatial correlation of the electric field is calculated in free space between locations $\mathbf{r}_1$ and $\mathbf{r}_2$ located in the *x-z* plane ($\Delta y = 0$) and at a distance $\Delta z = 0.05\lambda$.

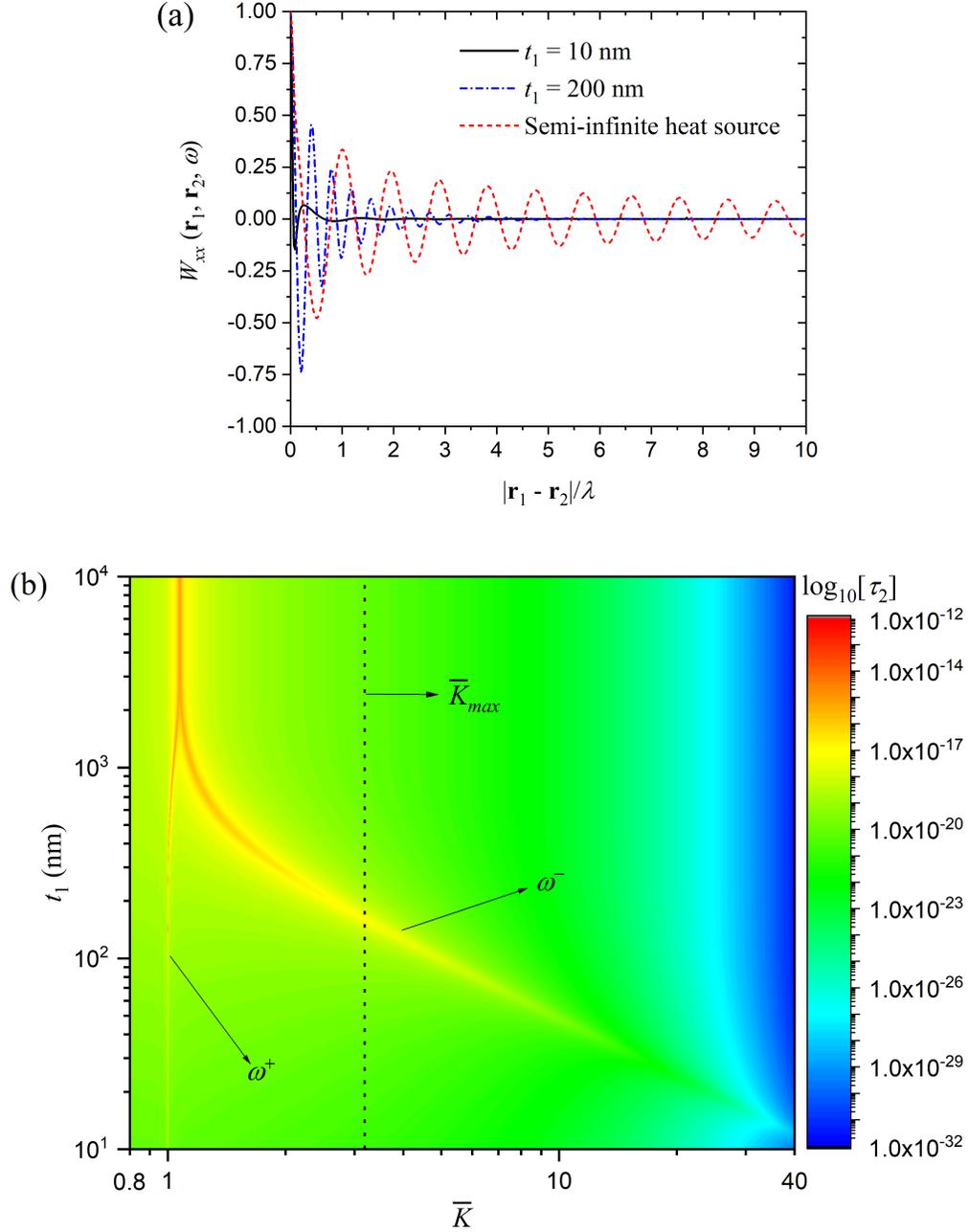

FIG. 4. (a) Element $W_{xx}$ of the cross-spectral density tensor for 10-nm and 200-nm-thick SiC heat sources surrounded by vacuum as a function of the length $|\mathbf{r}_1 - \mathbf{r}_2|$, normalized by the wavelength $\lambda$ (11.36 μm), calculated at a distance $\Delta z = 0.05\lambda$ in free space. $W_{xx}$ is normalized by its value at $|\mathbf{r}_1 - \mathbf{r}_2| = 0$. The results are compared against those for a semi-infinite SiC heat source. (b) TM-polarized transmission function, $\tau_2$, as a function of the heat source thickness, $t_1$, and the normalized parallel wavevector, $\overline{K}$, for a SiC source surrounded by vacuum. The frequencies $\omega^+$ and $\omega^-$ respectively refer to the anti-symmetric and symmetric modes.

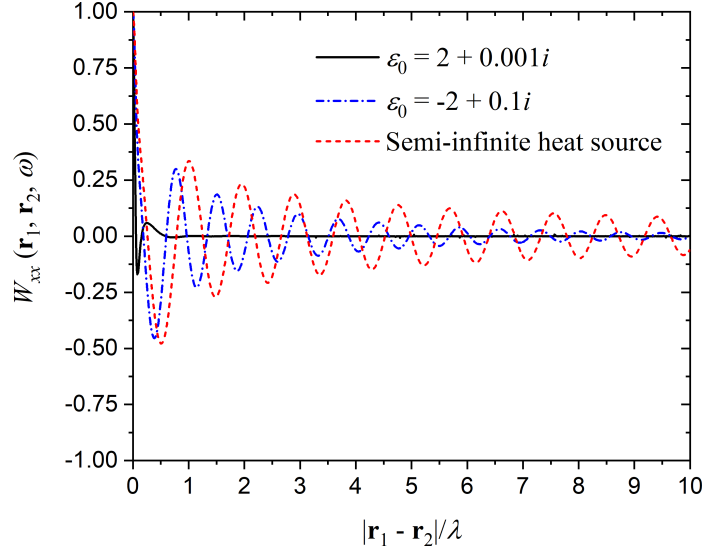

FIG. 5. Element $W_{xx}$ of the cross-spectral density tensor for a 10-nm-thick SiC heat source coated on dielectric ($\varepsilon_0 = 2 + 0.001i$) and metallic ($\varepsilon_0 = -2 + 0.1i$) substrates as a function of the length $|\mathbf{r}_1 - \mathbf{r}_2|$, normalized by the wavelength $\lambda$ (11.36 μm), calculated at a distance $\Delta z = 0.05\lambda$ in free space. $W_{xx}$ is normalized by its value at $|\mathbf{r}_1 - \mathbf{r}_2| = 0$. The results are compared against those for a semi-infinite SiC heat source.

Supplemental Material for article "Spatial correlation of the thermally generated electromagnetic field in layered media"

Vahid Hatamipour and Mathieu Francoeur[†]

*Radiative Energy Transfer Lab, Department of Mechanical Engineering, University of Utah, Salt Lake City, UT 84112, USA*

(Dated: January 12, 2020)

---

[†] Corresponding author. Tel.: + 1 801 581 5721
Email addresses: mfrancoeur@mech.utah.edu, vahid.hatami@utah.edu

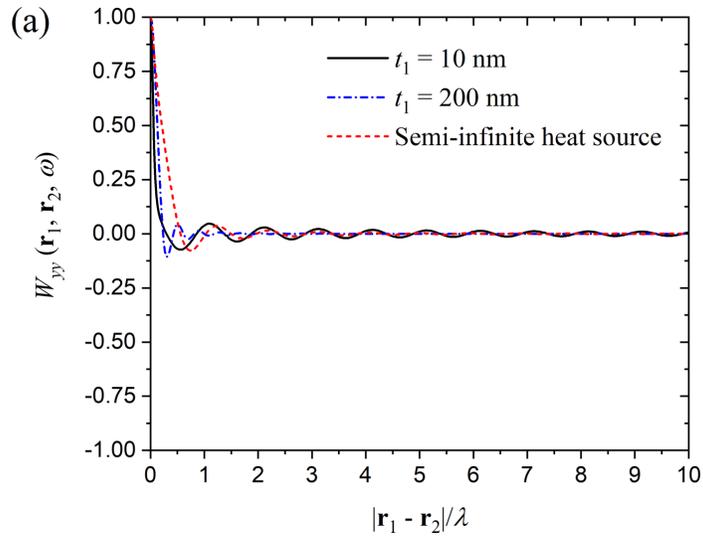

(a)

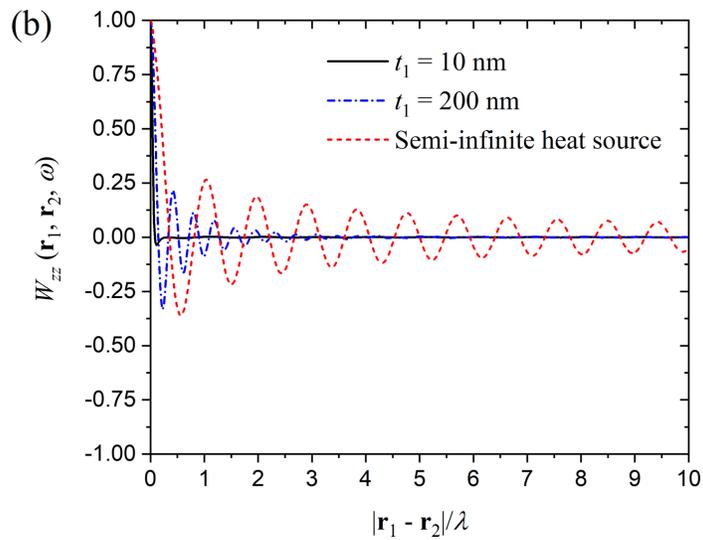

(b)

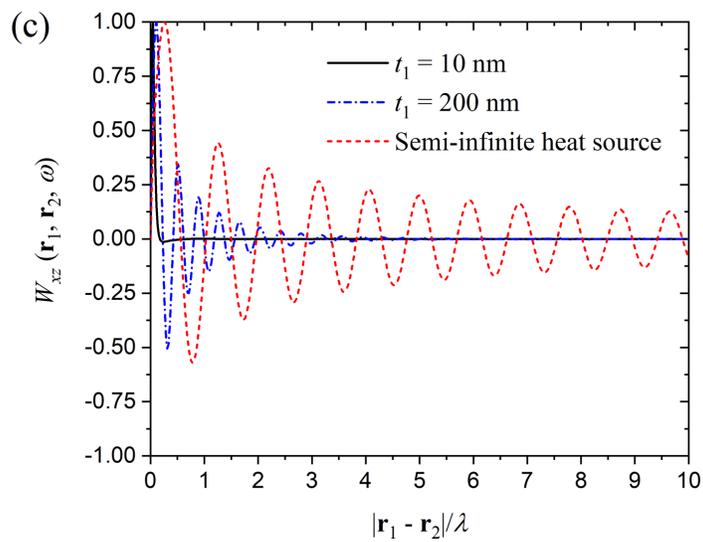

(c)

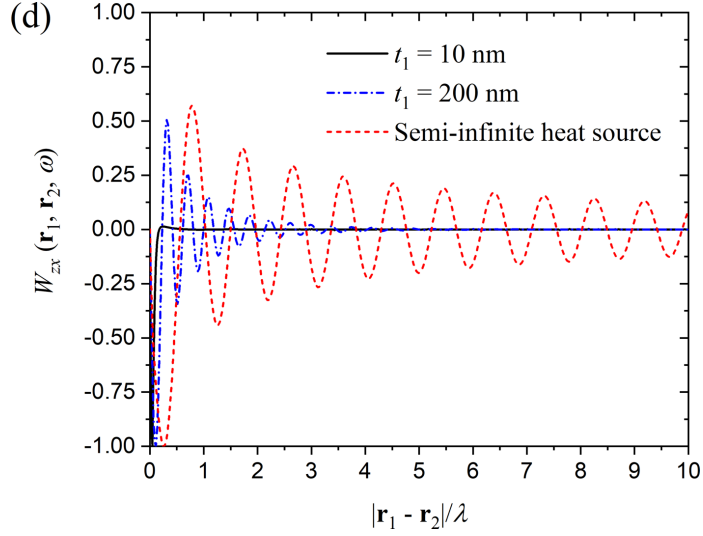

FIG. S1. Elements $W_{yy}$ (panel (a)), $W_{zz}$ (panel (b)), $W_{xz}$ (panel (c)) and $W_{zx}$ (panel (d)) of the cross-spectral density tensor for 10-nm and 200-nm-thick SiC heat sources surrounded by vacuum as a function of the length $|\mathbf{r}_1 - \mathbf{r}_2|$, normalized by the wavelength $\lambda$ (11.36 µm), calculated at a distance $\Delta z = 0.05\lambda$ in free space. $W_{yy}$, $W_{zz}$, $W_{xz}$ and $W_{zx}$ are normalized by their own values at $|\mathbf{r}_1 - \mathbf{r}_2| = 0$. The results are compared against those for a semi-infinite SiC heat source.

The elements $W_{zz}$, $W_{xz}$ and $W_{zx}$ are similar to $W_{xx}$. The spatial correlation length is maximum for a semi-infinite heat source and it decreases as the thickness of the heat source decreases. The element $W_{yy}$ results in a short spatial correlation length regardless of the thickness of the heat source. This is explained by the fact that SPhPs only exist in TM polarization (i.e., polarization in the $x$-$z$ plane) for SiC. Since SPhPs cannot be excited in TE polarization (i.e., polarization along the $y$-axis), the spatial correlation length for $W_{yy}$ is similar to that of a blackbody.

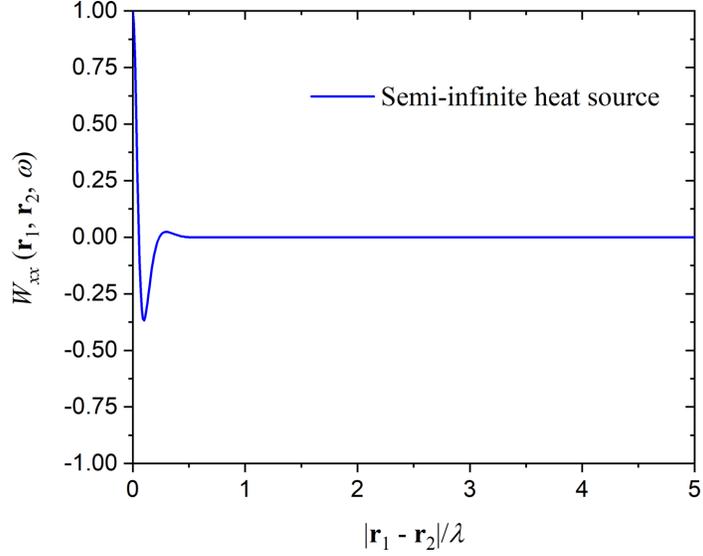

FIG. S2. Element $W_{xx}$ of the cross-spectral density tensor for a semi-infinite SiC heat source as a function of the length $|\mathbf{r}_1 - \mathbf{r}_2|$, normalized by the wavelength $\lambda$ (10.55 μm), calculated at a distance $\Delta z = 0.05\lambda$ in free space. $W_{xx}$ is normalized by its own value at $|\mathbf{r}_1 - \mathbf{r}_2| = 0$.

A semi-infinite SiC heat source is not spatially coherent at a wavelength of 10.55 μm which corresponds to the resonance of a single SiC-vacuum interface (arising when the real part of the dielectric function of SiC equals -1). Near-field thermal emission by a semi-infinite SiC heat source is quasi-monochromatic at a wavelength of 10.55 μm, which implies that a large number of $K$-modes are excited around that wavelength. Spatial coherence implies thermal emission within a narrow $K$-band, such that the spatial coherence length of a semi-infinite SiC heat source at 10.55 μm is similar to that of a blackbody.